\documentclass[%
twocolumn,
 amsmath,amssymb,
 aps, physrev,
]{revtex4-2}

\usepackage{graphicx}
\usepackage{dcolumn}
\usepackage{bm}
\usepackage{caption}
\usepackage{xcolor}

\begin{document}

\preprint{APS/123-QED}

\title{\textbf{Getting out of a tight spot:\\Cooperative unclogging of hydrogel particles in disordered porous media} 
}%

\author{Sanjana Kamath\textsuperscript{1}}
\author{Laurent Talon\textsuperscript{2}}
\email{Co-corresponding author:\\ laurent.talon@universite-paris-saclay.fr}
\author{Meera Ramaswamy\textsuperscript{1}}%
\author{Christopher A. Browne\textsuperscript{3}}
\author{Sujit S. Datta\textsuperscript{4,1}}
 \email{Corresponding author: ssdatta@caltech.edu}
\affiliation{\textsuperscript{1}Department of Chemical and Biological Engineering, Princeton University, Princeton, NJ 08544, USA}
\affiliation{\textsuperscript{3}Department of Chemical and Biomolecular Engineering, University of Pennsylvania, Philadelphia, PA 19104, USA}
\affiliation{\textsuperscript{2}Universit\'{e} Paris-Saclay, CNRS, FAST, 91405, Orsay, France.}
\affiliation{\textsuperscript{4}Division of Chemistry and Chemical Engineering, California Institute of Technology, Pasadena, CA 91125, USA}

\date{\today}

\begin{abstract}
We use event-driven pore network modeling to study the transport of hydrogel particles through disordered porous media---a process that underlies diverse applications. By simulating particle advection, deformation, and clogging at the pore scale, we identify a dimensionless ``squeezing parameter'' that quantitatively predicts the depth to which particles penetrate into a given medium across diverse conditions. 
Our simulations also uncover a surprising cooperative effect: adding more particles enables them to penetrate deeper into the medium. 
This phenomenon arises because individual particles redirect fluid to adjacent throats, forcing nearby particles through tight pores that they would otherwise clog. 
Altogether, these results help to establish a quantitative framework that connects microscopic particle mechanics to macroscopic transport behavior.

\end{abstract}

\maketitle



\textbf{\emph{Introduction.}} Hydrogels are crosslinked networks of hydrophilic polymers that can absorb up to $\sim10^3$ times their dry weight in water while maintaining structural integrity. The transport of hydrogel particles through porous media underlies a broad range of processes in energy, industry, medicine, and sustainability.
For example, hydrogel particle suspensions are used to clog high-permeability zones in oil reservoirs, promoting fluid redirection and oil recovery from bypassed lower-permeability zones~\cite{bai2007preformed,cao2020polymeric,chauveteau2003depth,yujun2003characteristics,pi2023visualized,lane2000gel,coste2000depth,tessarolli2018hydrogels}. A similar approach can improve the efficiency of geothermal energy extraction, which is similarly plagued by reservoir heterogeneity~\cite{doe2014discrete,winterfeld2025using}. In other applications such as groundwater remediation~\cite{hastings2021optimization,yang2023characteristics,wu2023novel} and drug delivery~\cite{li2016designing}, hydrogel particles are used to carry useful compounds, such as treatment chemicals and therapeutic agents, through porous groundwater aquifers and biological tissues/gels, respectively. All these uses of hydrogel particles require their spatial distribution through a disordered porous medium to be predictable and precisely controlled. However, achieving such predictability is still an open challenge; as a result, applications often proceed by trial and error, with highly variable results. Here, we address this challenge. 

The opacity of porous media typically precludes direct visualization of particle transport in situ.  Therefore, recent work has employed numerical simulations to characterize the transport of hydrogel particles in simplified settings. For example, studies have examined how hydrodynamic forces can enable a \emph{single} hydrogel particle to squeeze through an individual pore constriction (a ``throat'')~\cite{zhu2007modeling,fiddes2009augmenting,haghgooie2010squishy,merkel2011using,li2015universal}; more recent work extended these studies to the case of a deformable particle moving through a network of multiple interconnected pores~\cite{benet2018mechanical}. Other work hints that interactions between \emph{multiple} particles---both due to short-ranged collisions~\cite{o2019cooperative,shen2023anomalous} and longer-ranged alterations in the flow of the suspending fluid as particles squeeze into/clog pores~\cite{li2021modeling}---can strongly influence particle transport. However, these studies focused on ordered networks of identical pores, thus neglecting an inherent feature of most porous media: disorder arising from variations in the pore sizes, which can greatly alter flow and transport~\cite{datta2013spatial,wang2019disorder}. As a result, for a suspension of hydrogel particles of a given size, internal permeability, and stiffness, injected into a disordered porous medium of a given geometry, it is still not possible to predict what the spatial distribution of particles and resulting pore-scale changes in fluid flow will be.

\begin{figure*}[t]
    \centering
    \includegraphics[width=0.85\linewidth, trim = 10 65 115 205, clip=true]{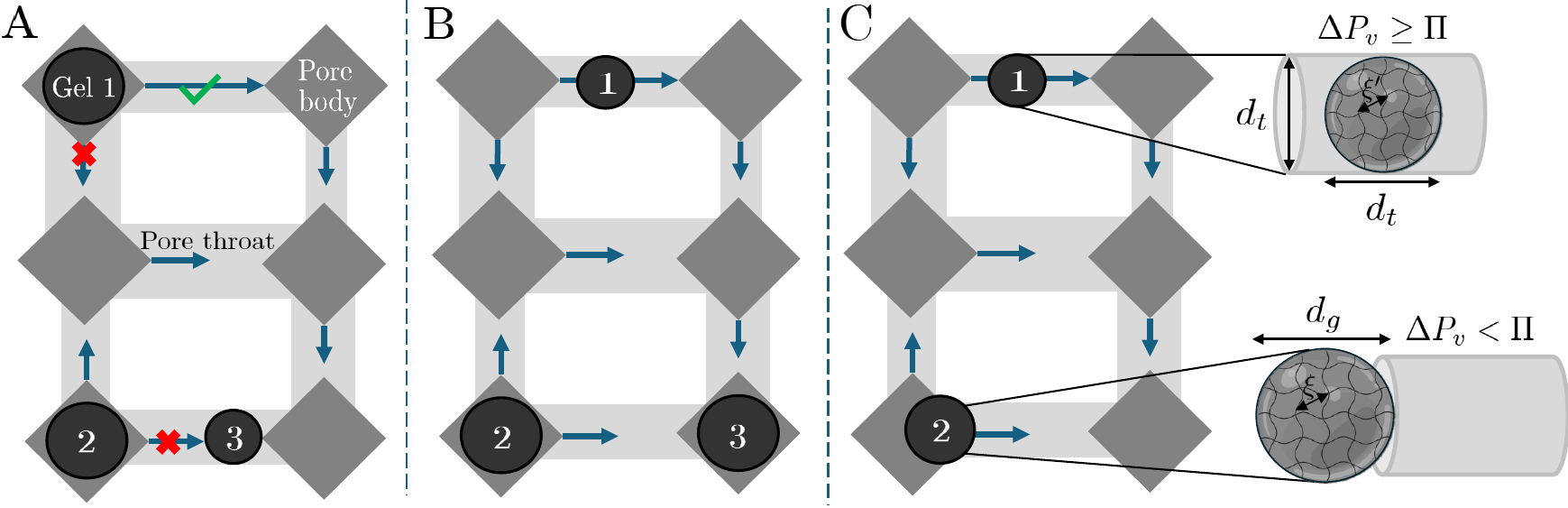}
    \caption{Schematic of the event-driven pore network model. (A) Hydrogel particles only move into vacant throats with the fastest local flow speed. (B) The particle with the shortest transit time across its throat (here \#3) moves into the next pore body while other particles move partially through their throats based on this transit time. (C) Hydrogel particles can only compress and squeeze into a pore throat if the viscous stress $\Delta P_v$ exceeds the compressive stress $\Pi$ (\#1); otherwise, they continue to clog the throat (\#2). Here, $d_g$ and $\xi$ represent the diameter and internal mesh size of the unconfined particle, while $\xi'$ represents the smaller mesh size of a compressed particle.}
    \label{fig:SimulationFramework}
\end{figure*}

Here, we address this gap in knowledge using event-driven pore network modeling of hydrogel particle suspensions injected into porous media of tunable degrees of disorder. We use simulations to characterize how the distance that particles penetrate into a medium varies with the imposed flow rate, degree of disorder in the medium, and hydrogel properties. Across the broad range of conditions tested, our results can all be quantitatively rationalized by considering the competition between the flow-induced viscous stress on a particle to the mechanical stress required to squeeze it through a pore throat. Our simulations also reveal a surprising phenomenon: Increasing the concentration of injected particles \emph{promotes} their penetration into a disordered medium, overturning the common assumption that more concentrated suspensions are more likely to cause clogging. We trace the origin of this cooperativity to dipolar alterations in the flow through the medium induced by individual particles. Altogether, our work establishes a quantitative framework to help predict and guide ways to control the distribution of hydrogel particles in a porous medium, with implications for diverse applications. 

\textbf{\emph{Event-driven pore network model.}} We follow the established approach of pore network modeling~\cite{fatt1956network,blunt2001flow,xiong2016review} by representing each porous medium as a one- or two-dimensional (1D or 2D) square lattice of pore throats (the edges of the network) connecting larger pore ``bodies'' (the nodes). This approach greatly simplifies the complexity of typical porous media while incorporating the essential physics of fluid flow through the different pores, flow-induced particle advection and squeezing into/clogging of individual pore throats, and the resulting changes in flow behavior and the distribution of particles over larger scales (Figure~\ref{fig:SimulationFramework}). 

In particular, we represent the pore throats as cylinders with diameters $d_t$ chosen randomly from a normal distribution with a prescribed standard deviation $\sigma$, which we refer to as the degree of disorder; we use the absolute value of any nonphysical negative values of $d_t$ if they are chosen. The length $l_t$ of each cylinder is then set based on the lattice spacing and diameters of the adjacent throats. We consider slow, viscous, incompressible flows with fluid injected into the throats spanning one side of the network (the ``inlet'') with a prescribed total flow rate $Q$. The pressure drop across a pristine throat is given by Poiseuille’s Law~\cite{li2021modeling,shen2023anomalous,benet2018mechanical}: $\Delta P_t= \frac{8\mu l_t}{\pi (d_t/2)^4}Q_t$, where $\mu$ is the fluid viscosity and $Q_t$ is the throat flow rate. In this representation, the advection, squeezing, and clogging of hydrogel particles is dictated solely by the geometry of the pore throats, which can only contain one compressed particle at a time; the bodies simply act as nodes in this network that can also only contain a single fully-swollen particle at a time, and their sizes are otherwise not explicitly considered. Future extensions of our work could relax this assumption by e.g., allowing multiple particles to enter pore bodies. Applying conservation of mass at each body then results in a linear system of equations that provides the quasi-steady-state flow profile of the network for each configuration of hydrogel particles, where the equilibration timescales for pressure and viscous flows are much shorter than the timescale of one particle moving through a throat. 

In our simulations, we combine this pore network flow model with event-driven hydrogel particle dynamics (Figure~\ref{fig:SimulationFramework})---numerically solving this system of equations successively as the distribution of particles throughout the network changes step-by-step. We initialize the particles in pore bodies along the inlet of the medium (Fig.~\ref{fig:SimulationFramework}A). The particles have a fully swollen diameter $d_{g}>\bar{d}_t$ and internal swollen mesh size $\xi$ (magnified schematic in the lower right of Fig.~\ref{fig:SimulationFramework}C), which then sets the internal permeability $\sim \xi ^2$ and bulk modulus $K\sim k_B T/ \xi ^3$~\cite{rubinstein2014polymer}. At each step of the simulation, we determine which hydrogel particles have vacant throats to potentially move into, advected by the flow out of the pore bodies they occupy. We then specify that each such particle will either squeeze into the accessible vacant throat that has the largest pristine flow speed (arrows in Fig.~\ref{fig:SimulationFramework}A--B, upper right of Fig.~\ref{fig:SimulationFramework}C) or become stuck and clog the throat entrance (lower right of Fig.~\ref{fig:SimulationFramework}C), depending on the local flow speed. 

In particular, the pressure drop across the throat is then given by $\Delta P_t+RQ_t$, where $R$ is known as the hydraulic resistance of the hydrogel particle and $Q_t$ is the reduced flow rate through both the particle and throat. For a fully swollen particle, $R\sim\mu/\left(d_g \xi^2\right)$~\cite{li2015universal}. However, the fluid flow through the particle compresses it~\cite{parker1987steady,quinn2013flow,li2015universal,hewitt2016flow,macminn2016large,xu2024hystereses}. Assuming isotropic compression for simplicity, the compressed diameter is $d_g'<d_g$, the mesh size is then $\xi'=\xi\left(d_g'/d_g\right)^{9/4}$~\cite{de1979scaling,gao2021scaling}, the hydrogel resistance is $R\sim \mu d_g^{\prime}/\left(d_t \xi^{\prime}\right)^2$~\cite{li2015universal}, and the bulk modulus is $K\sim k_BT/\xi'^3$~\cite{rubinstein2014polymer}; investigating the influence of anisotropic deformation mechanics will be an interesting extension of our work. So, in this step of the simulation, we determine the diameter of the compressed particle by balancing the viscous pressure drop across the particle, $RQ_t$, with the mechanical stress required to compress the hydrogel, $\Pi=K\varepsilon$, where $\varepsilon\equiv\left(d_g^3-d_g'^3\right)/d_g^3$~\cite{li2015universal}, and then update all flow rates throughout the network accordingly. If $d_g'>d_t$, then the particle cannot fully squeeze into the throat and remains stuck in the upstream pore body, clogging the throat entrance. Once the flow is sufficiently fast to reach the threshold $d_g'=d_t$, however, the particle is advected through the throat (upper right of Fig.~\ref{fig:SimulationFramework}C).

Having thereby identified which hydrogel particles will squeeze through throats, we then determine how far they will be advected during this step of the simulation.  To do so, we determine the total time required for each particle with $d_g'=d_t$ to transit through its pore throat, based on the local flow speed. The particle with the shortest transit time---which sets the duration of this step of the simulation---is then advected to its downstream pore body, while the others move only partially through their respective throats as dictated by the local flow speed (Fig.~\ref{fig:SimulationFramework}B). Our numerical simulation then repeats this series of steps---which we refer to as ``event-driven'' because the temporal duration of each step is not fixed, but is determined by the dynamics of the fastest-moving particle in the network---until all particles are stuck ($d_g'>d_t$) or have exited the medium.

\textbf{\emph{Single particle dynamics.}} To establish a simple baseline, we first examine the transport of a single hydrogel particle across a range of inlet flow rates $Q$ through a 1D medium. For each value of $Q$ tested, we simulate $10^3$ different but statistically identical media with the same mean diameter $\bar{d}_t=0.8d_g$ and prescribed degree of disorder $\sigma$. In each simulation, we measure the number of throats $x$ that the particle penetrates along the length of the medium before it can no longer move. 

\begin{figure}[b]
    \centering
    \includegraphics[width=1\linewidth,trim = 100 225 100 235, clip=true]{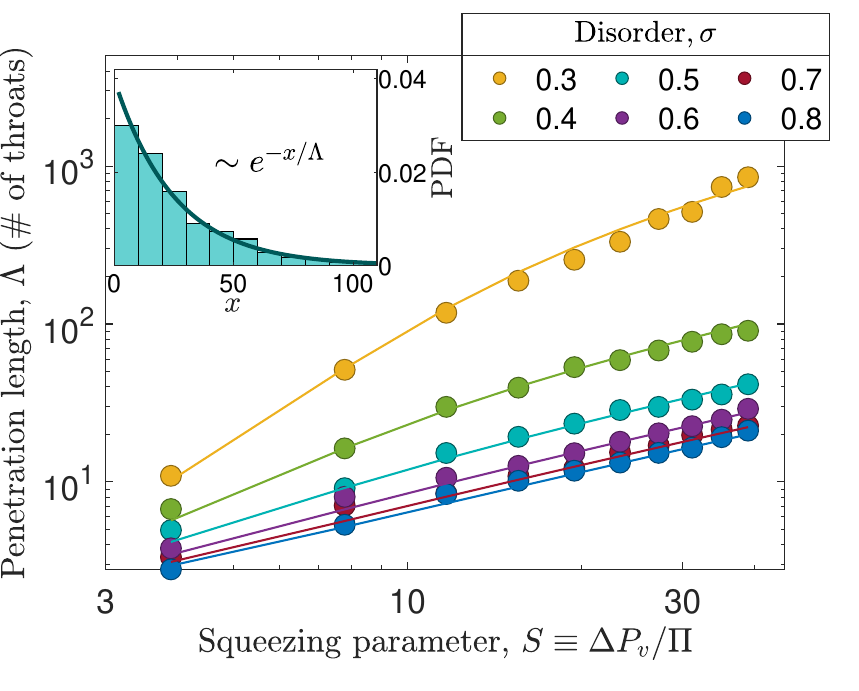}
    \caption{Penetration of hydrogel particles into 1D porous media. Inset shows the probability density function of the number of throats $x$ that the particle penetrates into the medium for 1000 simulations of a medium with $\sigma=0.5$ and $S \approx 12$, following the exponential distribution with mean value $\Lambda$ as shown by the curve. Main panel shows $\Lambda$ increasing with increasing squeezing parameter $S$ and decreasing degree of disorder $\sigma$; curves show the theoretical prediction of Eq.~\eqref{penetration}. All simulations use $\mu = 10^{-3}~\mathrm{Pa}\cdot\mathrm{s}$, $\bar{d}_t=1~\mathrm{mm}$, $d_{g}=1.2\bar{d}_t$, and $\xi=10^{-8}~\mathrm{m}$ in media composed of 200 successive throats, with the exception of $\sigma = 0.3$ that is instead composed of 1000 throats to accommodate for deeper penetration at low disorder.}
    \label{fig:1Gelin1D}
\end{figure}

The probability density function (PDF) of $x$ determined from all simulations of the exemplary case of $\sigma=0.5$ and $Q=3\times 10^{-12} ~\mathrm{m^3/s}$ is given in the inset of Fig.~\ref{fig:1Gelin1D}. As shown by the curve, the $\mathrm{PDF}\sim e^{-x/\Lambda}$, where we term $\Lambda$ the mean penetration length (normalized by $\bar{d}_t$). As intuitively expected, our simulations indicate that $\Lambda$ increases monotonically with increasing $Q$: hydrogel particles penetrate deeper into the medium at higher flow rates. This intuition can be quantified by comparing the characteristic fluid viscous stress on a compressed particle $\Delta P_v\sim\mu Q_t/\left(d_g'\xi'^2\right)$ and the characteristic compressive stress $\Pi\sim K\varepsilon$, with $d_g'={d}_t$ and $Q_t=Q$ in 1D; the ratio between these stresses yields a dimensionless group that we term the ``squeezing parameter'', $S\equiv\Delta P_v/\Pi$. As shown by the light blue points in Fig.~\ref{fig:1Gelin1D}, $\Lambda$ increases as $S$ increases above a critical value $\sim1$, as expected. Moreover, repeating our simulations for a given degree of disorder $\sigma$ but independently varying the other input parameters $\bar{d}_t$, $\xi$, and $\mu$ instead of $Q$ yields values of $\Lambda$ that collapse when presented as a function of $S$ (\emph{Supplementary Material}), confirming that this dimensionless parameter controls the extent of particle penetration.

This picture suggests that the probability of a hydrogel particle getting stuck in a 1D medium is independent of space and time~\cite{vankampen}---as further supported by our finding that the PDF of penetration lengths decays exponentially, indicative of a Poisson process. Specifically, for a given set of input conditions parameterized by $S$, the critical throat diameter $d_t^*(S)$ for particle squeezing is given by equating $\Delta P_v$ to $\Pi$. As detailed in the \emph{Supplementary Material}, the probability that the particle gets stuck is then equal to the probability that it encounters a throat with $d_t<d_t^*$, and $\Lambda$ is simply the inverse of this probability: 
\begin{equation}
\label{penetration}
\Lambda(S) = \mathrm{ln}\left[ 1- \frac{1}{2} \mathrm{erf}\left( \frac{d_t^*(S)-\bar{d}_t}{\sigma \sqrt{2}} \right) + \mathrm{erf}\left( \frac{d_t^*(S)+\bar{d}_t}{\sigma \sqrt{2}} \right) \right]^{-1}.
\end{equation}
As shown by the light blue curve in Fig.~\ref{fig:1Gelin1D}, this prediction agrees well with the simulation data, confirming our intuitive picture. As a final test of this picture, we repeat our simulations for media with varying degrees of disorder $\sigma$. In all cases, $\Lambda$ increases monotonically with $S$ in excellent agreement with Eq.~\eqref{penetration}, as shown by the different colors in Fig.~\ref{fig:1Gelin1D}. These results demonstrate that in 1D, disorder results in a higher likelihood of the particle encountering a throat that is too small for it to pass through, hindering penetration into the medium.

\begin{figure}[t]
    \centering
    \includegraphics[width=1\linewidth,trim = 100 225 95 240, clip=true]{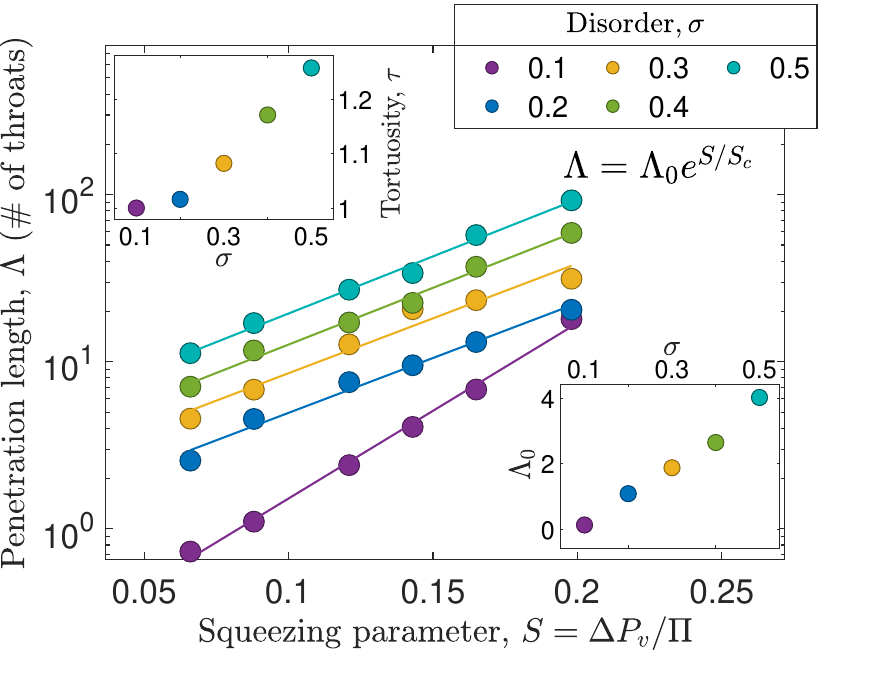}
    \caption{Penetration of hydrogel particles into 2D porous media. Main panel shows $\Lambda$ increasing with increasing squeezing parameter $S$ and degree of disorder $\sigma$; curves show exponential fits $\Lambda_0e^{S/S_c}$. Top left and bottom right insets show that both the particle path tortuosity $\tau$ and characteristic penetration length $\Lambda_0$ increase with disorder $\sigma$. All simulations use $\mu = 10^{-3}~\mathrm{Pa}\cdot\mathrm{s}$, $\bar{d}_t=1~\mathrm{mm}$, $d_{g}=1.2\bar{d}_t$, and $\xi=10^{-8}~\mathrm{m}$ in media composed of $120\times120$ throats.}    
    \label{fig:1Gelin2D}
\end{figure}

Intriguingly, this trend is reversed in 2D: as shown in Fig.~\ref{fig:1Gelin2D}, for a given value of $S$, increasing the degree of disorder $\sigma$ \emph{increases} the hydrogel particle penetration length $\Lambda$. Moreover, while $\Lambda$ again increases monotonically with $S$, it does so exponentially --- not as predicted by Eq.~\eqref{penetration} for the 1D case, nor as a power law as expected from percolation theory~\cite{benet2018mechanical}--- as shown by the lines of $\Lambda=\Lambda_0 e^{S/S_c}$ in Fig.~\ref{fig:1Gelin2D} with fitting parameters $\Lambda_0$ and $S_c$. The lower right inset shows how $\Lambda_0$ monotonically increases with $\sigma$, again reflecting that particle penetration is promoted by pore space disorder. Both of these differences with the 1D case result from the increased connectivity of the pore space in 2D: the latter provides an additional degree of freedom for particle squeezing and advection through throats with the fastest flow speeds, which follows tortuous paths through the medium that avoid the smallest throats~\cite{benet2018mechanical}. As shown in the upper left inset of Fig.~\ref{fig:1Gelin2D}, our simulations confirm that the tortuosity $\tau$ of these paths increases with the degree of disorder $\sigma$. Moreover, the critical value of $S$ required for particle penetration is approximately two orders of magnitude lower than in the 1D case---again due to the increased pore space connectivity (\emph{Supplementary Material}), since fluid can now route through the $N$ throats transverse to a clogged one, thereby reducing its flow rate by a factor of $N$ (which is indeed $\sim10^2$ in the simulations).

\begin{figure}
    \centering
    \includegraphics[width=1\linewidth,trim = 95 70 130 185, clip=true]{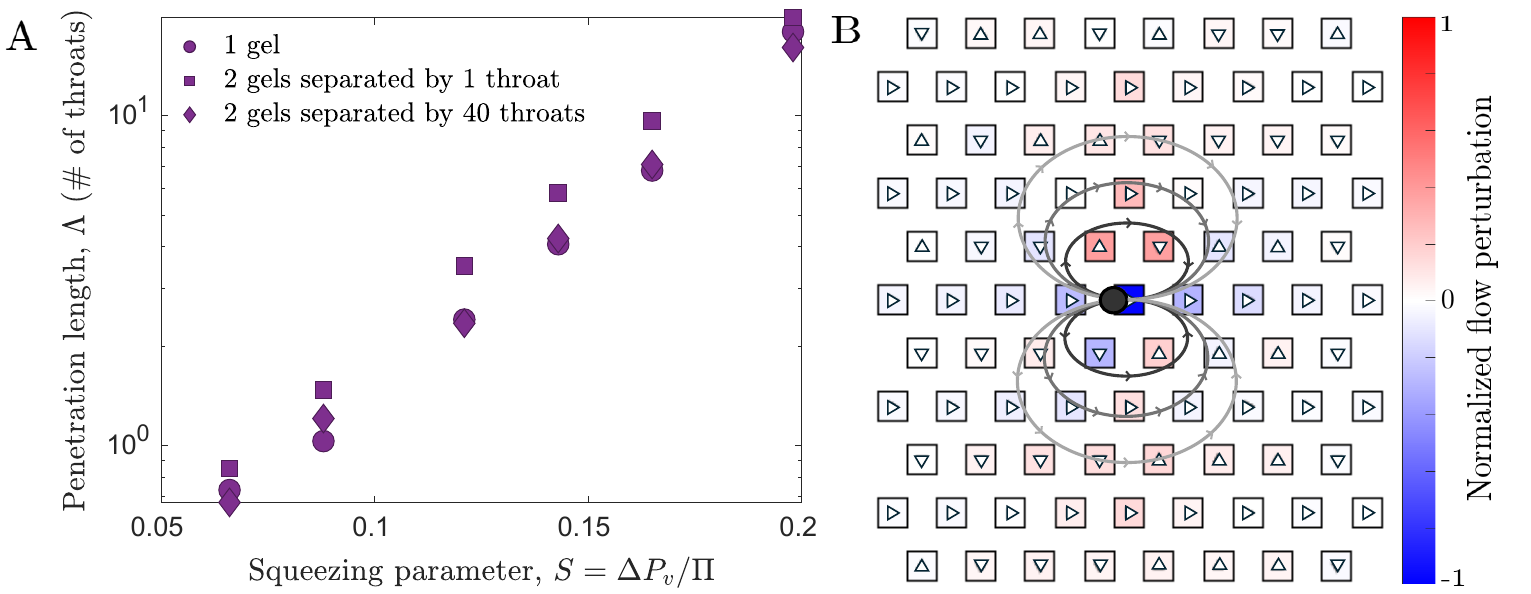}
    \caption{Cooperative unclogging of hydrogel particles is mediated by dipolar flow perturbations. (A) Mean penetration length $\Lambda$ as a function of $S$ for a single particle (circles), 2 particles separated transversely by a single throat (squares), or 2 particles separated transversely by 40 throats (diamonds), in the same 2D medium with $\sigma=0.1$ and $120\times120$ throats. The second case has $\approx30\%$ more penetration. (B) Dipolar perturbation to the flow induced by a particle clogging a pore. The color bar shows the relative change in the flow speed caused by a hydrogel particle; sketched streamlines show the dipolar nature of the flow.}
    \label{fig:2Gels}
\end{figure}

\textbf{\emph{Cooperative dynamics of multiple particles.}} Next, to probe how multiple hydrogel particles are transported, we repeat our 2D simulations with two particles at different initial transverse separation distances from each other. Remarkably, when both particles are initially next to each other (separated by one throat), the mean penetration length is $\approx30\%$ larger than in the single-particle case, as shown by comparing the squares to circles in Fig.~\ref{fig:2Gels}A. This effect diminishes as the initial transverse separation between particles increases, as we show in the \emph{Supplementary Material}; as a limiting case, when the particles are separated by as many as 40 pore throats, their penetration into the medium is similar to the single-particle case, as shown by the diamonds in Fig.~\ref{fig:2Gels}A. Alternatively, this unexpected effect is promoted as more hydrogel particles are added to the medium, as we also show in the \emph{Supplementary Material}. This cooperative effect is also sensitive to the relative orientation of the particles: penetration is only promoted when the particles are close to each other in the transverse, not longitudinal direction (\emph{Supplementary Material}). Taken altogether, these results indicate that the cooperative unclogging of multiple hydrogel particles arises from orientation-dependent alterations in the flow through the medium induced by individual particles. Close inspection of the fluid velocities around a single stuck hydrogel particle confirms and elucidates this expectation. As shown in Fig.~\ref{fig:2Gels}B, each particle induces a dipolar perturbation to the fluid velocity field characteristic of low Reynolds number flow around an obstacle~\cite{champagne2010traffic,gavrilchenko2019resilience}. Thus, when one particle clogs or squeezes through a pore, it redirects fluid to neighboring pores, enabling nearby particles to also squeeze through tight pores.

\textbf{\emph{Discussion.}} Our work demonstrates that the penetration of hydrogel particles into a porous medium is governed by the competition between the flow-induced viscous stress on a particle and the mechanical stress required to squeeze it through constrictions---providing a quantitative foundation for their applications in energy, industry, medicine, and sustainability. Our simulations highlight the pivotal role of dimensionality and disorder: in 1D, increasing disorder hinders particle penetration, whereas in higher-dimensional media, greater disorder actually promotes deeper penetration due to the availability of alternative flow paths. Moreover, the simulations reveal the counterintuitive finding that increasing the particle concentration enhances penetration rather than promoting clogging. This cooperative unclogging occurs through dipolar alterations in the local flow field, where particles redirect fluid to neighboring pores, facilitating the transport of nearby particles through tight constrictions. Exploring how our findings extend to impermeable but still deformable particles, such as cells, vesicles ~\cite{rogers2023characterizing}, emulsion droplets ~\cite{soo1984flow}, and colloidal aggregates~\cite{bizmark2020multiscale}, will be a useful direction for future work. \\

\textbf{\emph{Acknowledgments.}} It is a pleasure to acknowledge Baojun Bai for useful discussions, as well as funding to SK through the Chateaubriand Fellowship of the Office for Science \&
Technology of the Embassy of France in the United States, to LT through the INTPART program from the Research Council of Norway (project number 309139), and to SSD through NSF grant DMR-2011750, the Project X Innovation Fund, the Camille Dreyfus Teacher-Scholar Program of the Camille and Henry Dreyfus Foundation
(SSD). This material is also based upon work by SSD supported by the U.S. Department of Energy’s Office of Energy Efficiency and Renewable Energy (EERE) under the Geothermal Technologies Office (GTO) Innovative Methods to Control Hydraulic Properties of Enhanced Geothermal Systems Award Number DE-EE0009790.

\newpage\section*{Supplementary Material}

\textbf{\emph{Clogging of a single particle in 1D.}} For a given inlet flow rate $Q$, which  is constant through each of the throats in the medium, there exists a critical throat diameter $d_t^*(Q)$ for gel particle squeezing: particles can successfully squeeze through throats with $d_t\geq d_t^*(Q)$ and instead clog throats with $d_t< d_t^*(Q)$. This critical diameter is given by the balance between the viscous pressure drop across a particle compressed to the same size as $d_t^*$, $\Delta P_v=Q\mu /\left(d_t^* \xi'^2\right)$, and the compressive stress, $\Pi=K\varepsilon$, where $\xi'=\xi\left(d_t^*/d_g\right)^{9/4}$, $K\sim k_BT/\xi'^3$, and $\varepsilon\equiv\left(d_g^3-d_t^{*3}\right)/d_g^3$. That is, we invert the following equation and solve for $d_t^*(Q)$:
\begin{equation}
Q = \frac{k_bTd_t^*}{\mu \left[ \xi(d_t^*/d_g)^{9/4} \right]}\frac{d_g^3-d_t^{*3}}{d_g^3}.
\end{equation}
Thus, in 1D, the probability $P$ that a hydrogel particle travels through $x$ number of throats before getting stuck is independent of position and time, and follows the discrete Poisson law:
\begin{equation}
\label{PoissonLaw}
P(x) = \alpha (1-\alpha)^{x},
\end{equation}
where $\alpha$ is the probability that the particle encounters a throat with $d_t<d_t^*$. We obtain this probability by taking the cumulative distribution function of the probability of obtaining a certain throat diameter, $p(d_t)$:
\begin{equation}
\label{Integral}
\alpha = P(d_t<d_t^*) = \int_{-d_t^*}^{d_t^*} p(d_t)\mathrm{d}d_t,
\end{equation}
\begin{equation}
\label{Integral2}
P(d_t<d_t^*) = \frac{1}{2} \mathrm{erf}\left( \frac{d_t^*-\bar{d_t}}{\sigma \sqrt{2}} \right) + \mathrm{erf}\left( \frac{d_t^*+\bar{d_t}}{\sigma \sqrt{2}} \right).
\end{equation}
The bounds of the integral range from $-d_t^*$ and $d_t^*$ to account for the fact that we have taken the absolute value of the random normal distribution when generating the pore throat diameters. With $P(x)=Ae^{-x/\Lambda}$, Eq.~\eqref{PoissonLaw} then directly yields $\Lambda = \left[ -\mathrm{ln}(1-\alpha)\right ]^{-1}$, where $\alpha$ is given by Eqs.~\eqref{Integral} and~\eqref{Integral2}, as presented in the main text.\\

\textbf{\emph{Derivation of the \emph{S} parameter.}} Consider two parallel throats in a 2D medium, one clogged by a hydrogel particle and the other not. Equating the fluid pressure drop across them, $\approx\frac{\mu Q_{t}^\prime d_{t}}{(d_{t} \xi^\prime)^2}$ and $\frac{8\mu Q_{t} l_{t}}{\pi (d_{t}/2)^4} $, respectively, then yields the flow rate through the clogged throat: $Q_{t}^\prime = \frac{128l_{t} \xi^{\prime ^2}}{\pi d_{t}^3}Q_{t}
$, where $Q_t$ is the flow rate through the unclogged throat, $l_t$ and $d_t$ are the throat length and diameter respectively, and $\xi'$ is the mesh size of the compressed hydrogel. If the medium spans $N$ throats across, then $Q = Q_t(N-1) + Q_{t}\frac{128l_{t} \xi^{\prime ^2}}{\pi d_{t}^3}$ and therefore $Q_{t} = \frac{Q}{(N-1)+ \frac{128l_{t}\xi^{\prime ^ 2}}{\pi d_{t}^3}}$. The viscous pressure drop across a throat is then given by
\begin{equation}
\Delta P_{v} = \frac{8\mu Q_{t}l_{t}}{\pi(d_{t}/2)^4}= \frac{\frac{8\mu Ql_t}{(N-1)+ \frac{128l_{t}\xi^{\prime ^ 2}}{\pi d_{t}^3}}}{\pi(d_{t}/2)^4},
\end{equation}
and the general expression for $S\equiv\frac{\Delta P_{v}}{K^{\prime}\frac{(d_g^3-d_t^3)}{d_g^3}}$ is:
\begin{equation}
S = \frac{\frac{8\mu Q d_t/[(N-1)+ \frac{128d_{t}\xi^{\prime ^ 2}}{\pi d_{t}^3}]}{\pi(d_{t}/2)^4}}{\frac{k_bT}{\xi^{\prime ^ 3}}\frac{(d_g^3-d_t^3)}{d_g^3}}
\label{ExtendedS}
\end{equation}
where we have substituted $K^{\prime}=\frac{k_bT}{\xi^{\prime ^ 3}}$ for the bulk modulus of the compressed gel and $l_t$ is set equal to $d_t$.

In the 1D case, $N=1$, and Eq.~\eqref{ExtendedS} simplifies to:
\begin{equation}
S = \frac{\frac{\mu Q}{d_t \xi^{\prime^2}}}{{\frac{k_bT}{\xi^{\prime ^ 3}}\frac{(d_g^3-d_t^3)}{d_g^3}}}.
\end{equation}
In the 2D case with $N\gg1$, we instead have:
\begin{equation}
S \approx\frac{\frac{8\mu d_t \frac{Q}{(N-1)}}{\pi(d_{t}/2)^4}}{\frac{k_bT}{\xi^{\prime ^ 3}}\frac{(d_g^3-d_t^3)}{d_g^3}}.
\end{equation}\\

\begin{figure}[h]
    \centering
    \includegraphics[width=1\linewidth,trim = 100 225 100 235, clip=true]{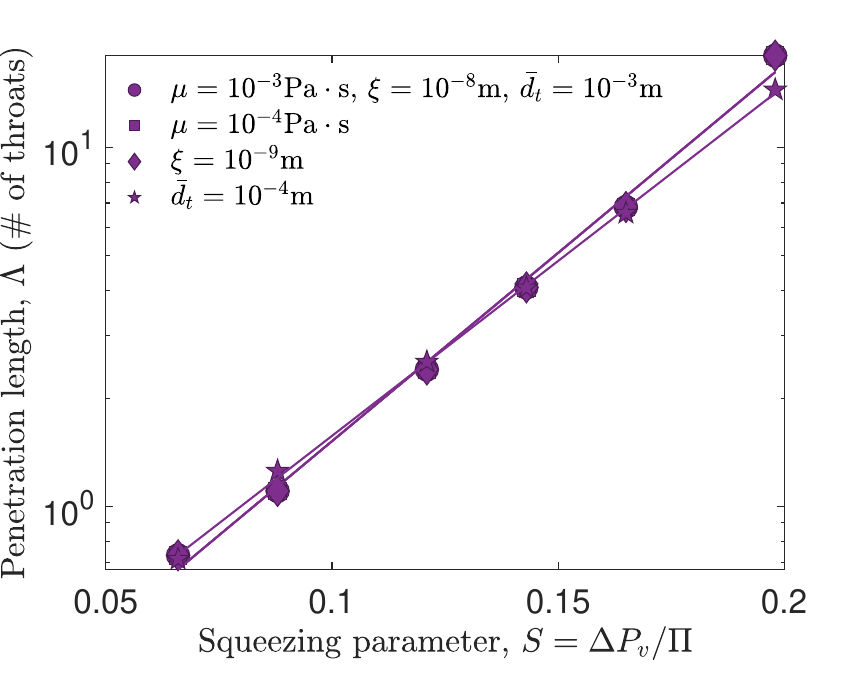}
    \caption{Mean penetration length $\Lambda$ as a function of $S$ for various fluid viscosities, hydrogel mesh sizes, and medium mean throat diameters. Medium size is $120\times120$ with $\sigma = 0.1$. The curve for $\bar{d_t}=10^{-4} ~\mathrm{m}$ is slightly offset from the rest because this curve was generated using a different set of media; the variation in the overall distribution of throat sizes leads to this minor discrepancy in the plot.
    }
    \label{fig:ViscMeshThroat}
\end{figure}

\begin{figure}[h]
    \centering
    \includegraphics[width=1\linewidth,trim = 100 225 100 235, clip=true]{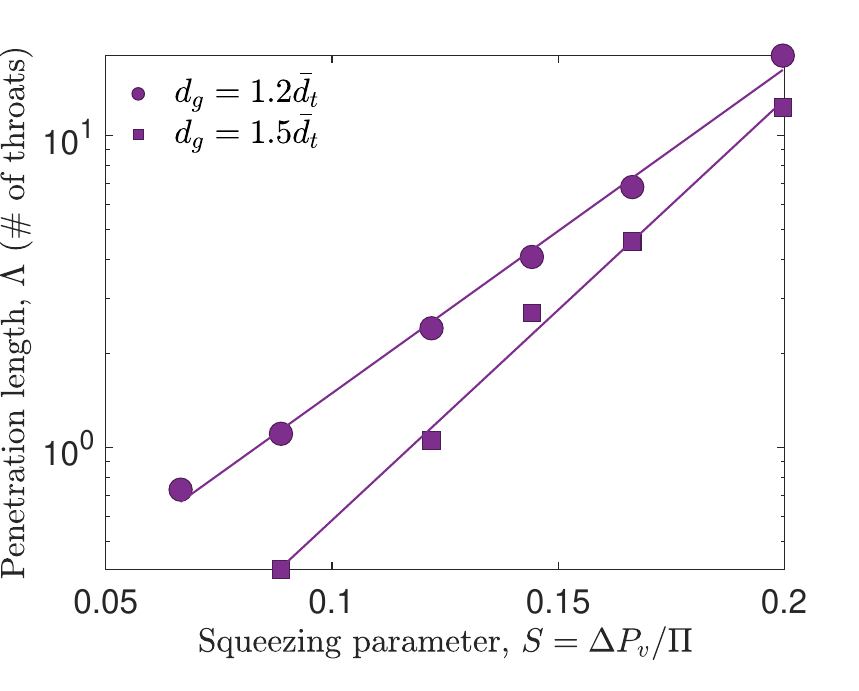}
    \caption{Penetration of hydrogel particles with different diameters in 2D porous media. All simulations use $\mu = 10^{-3}~\mathrm{Pa}\cdot\mathrm{s}$, $\bar{d}_t=1~\mathrm{mm}$, $\xi=10^{-8}~\mathrm{m}$, and $\sigma = 0.1$. Medium size is $120\times 120$.}
    \label{fig:GelSize}
\end{figure}

\begin{figure}[h]
    \centering
    \includegraphics[width=1\linewidth,trim = 100 225 100 235, clip=true]{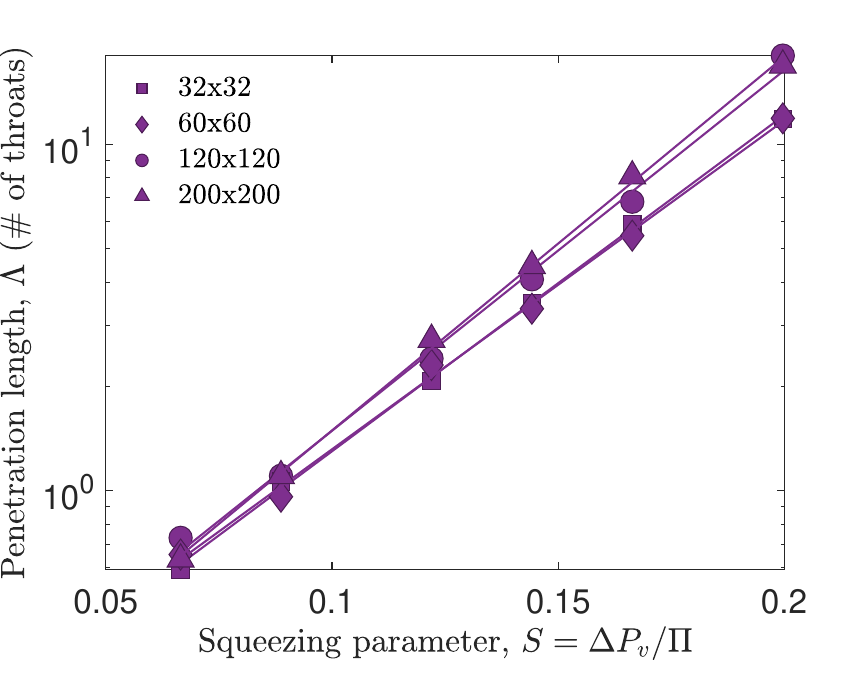}
    \caption{Mean penetration length $\Lambda$ as a function of $S$ for various system sizes. All simulations use $\mu = 10^{-3}~\mathrm{Pa}\cdot\mathrm{s}$, $\bar{d}_t=1~\mathrm{mm}$, $\xi=10^{-8}~\mathrm{m}$, and $\sigma = 0.1$.}
    \label{fig:SystemSize}
\end{figure}

\textbf{\emph{Influence of varying system parameters.}} The main text reports results of simulations testing varying inlet flow rates $Q$, keeping the other parameters constant: $\mu = 10^{-3}~\mathrm{Pa \cdot s}$, $\xi = 10^{-8} ~\mathrm{m}$, $\bar{d_t}=10^{-3}~\mathrm{m}$, and $d_g=1.2\bar{d}_t$, in 2D networks composed of $N\times N$ pore throats with $N=120$. Here, we present the results of simulations that test the influence of these other system parameters, varying each one at a time while maintaining the other parameters at the default values. As shown in Figure~\ref{fig:ViscMeshThroat}, all the data obtained for varying $\mu$, $\xi$, and $\bar{d}_t$ collapse on each other when presented as a function of the $S$ parameter. As shown in Fig.~\ref{fig:GelSize}, as expected, smaller particles are able to penetrate father into the medium than larger particles over the same range of flow rates (or $S$ values). Finally, as shown in Fig.~\ref{fig:SystemSize}, data obtained from simulations with varying number of pore throats $N$ converge for $N\geq120$, indicating that the other simulation results presented throughout the paper (which use $N=120$) are for a sufficiently large system. \\

\begin{figure}[h]
    \centering
    \includegraphics[width=1\linewidth,trim = 100 225 100 235, clip=true]{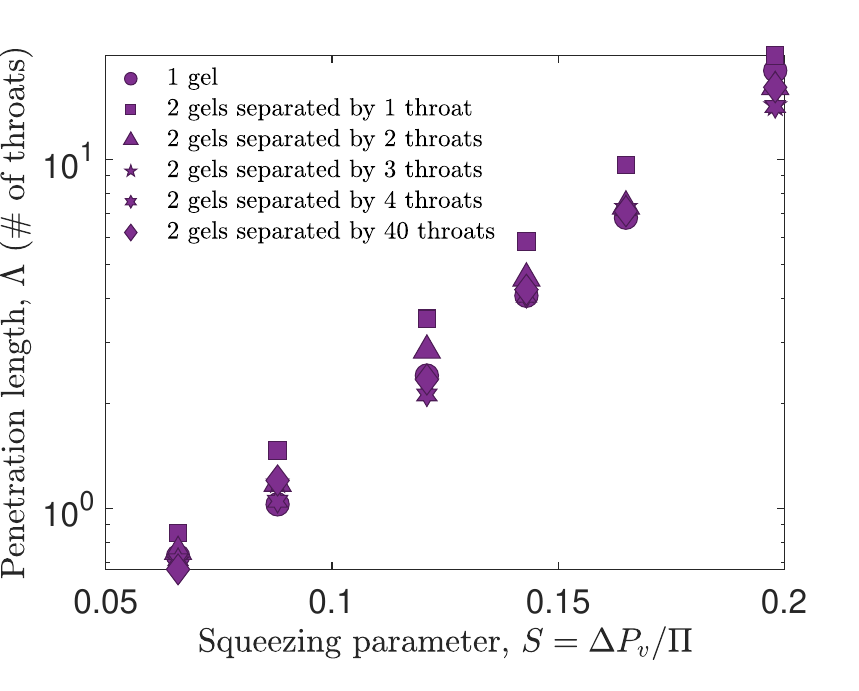}
    \caption{Mean penetration length $\Lambda$ as a function of $S$ for two particles with varying separation distances. All simulations use $\mu = 10^{-3}~\mathrm{Pa}\cdot\mathrm{s}$, $\bar{d}_t=1~\mathrm{mm}$, $\xi=10^{-8}~\mathrm{m}$, and $\sigma = 0.1$. Medium size is $120\times120$.}
    \label{fig:GelSeparation}
\end{figure}

\textbf{\emph{Two particles in different configurations.}} In this section, we examine how varying the placement of two hydrogel particles impacts their penetration length. 
First, we examine the influence of changing the separation distance between transverse-separated particles. As shown in Fig.~\ref{fig:GelSeparation}, we observe that penetration is enhanced most prominently when the particles are at a minimal separation distance (1 throat). The other curves representing increasingly large separation distances begin to approach one another, until they quickly (at a separation greater  than two throats) overlap with the single particle case.

\begin{figure}[h]
    \centering
    \includegraphics[width=1\linewidth,trim = 100 225 100 235, clip=true]{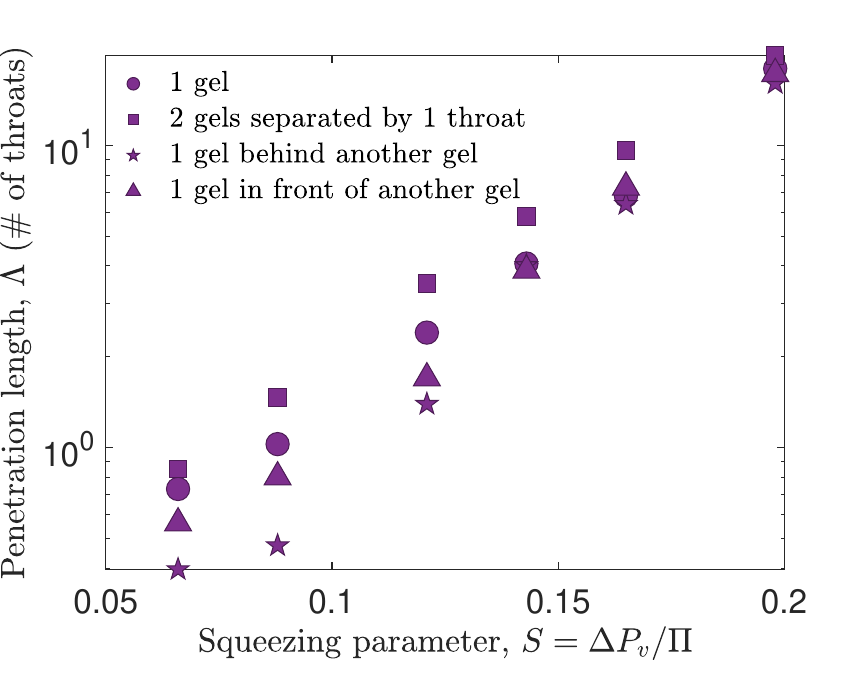}
    \caption{Mean penetration length $\Lambda$ as a function of $S$ for two particles placed in front of each other. All simulations use $\mu = 10^{-3}~\mathrm{Pa}\cdot\mathrm{s}$, $\bar{d}_t=1~\mathrm{mm}$, $\xi=10^{-8}~\mathrm{m}$, and $\sigma = 0.1$. Medium size is $120\times120$.}
    \label{fig:GelsinFront}
\end{figure}

Next, we examine the case where two particles are placed in front of one another (in the longitudinal direction). We compute the penetration length separately for each of the two particles since their behavior is distinct from one another. As shown in Fig.~\ref{fig:GelsinFront}, the penetration length is drastically reduced for the particle that is placed behind the leading particle, which is directly obstructing its path. In comparison, the particle that starts out in front has a larger penetration length. Still, each of the two particles travels a shorter distance than a single particle, while the two particles in the original configuration (separated by one throat in the transverse direction) travel the farthest. These data illustrate that the dipolar flow perturbation only enhances penetration when the particles are placed as close as possible in the transverse, rather than longitudinal, direction.

\begin{figure}[h]
    \centering
    \includegraphics[width=1\linewidth,trim = 100 225 100 235, clip=true]{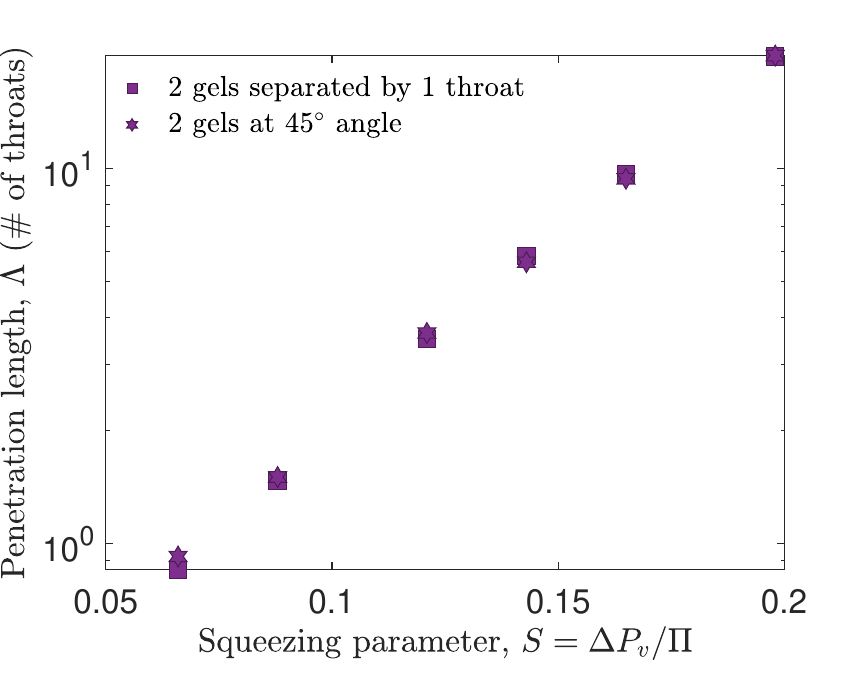}
    \caption{Mean penetration length $\Lambda$ as a function of $S$ for two particles initially placced at a $45^{\circ}$ angle from each other. All simulations use $\mu = 10^{-3}~\mathrm{Pa}\cdot\mathrm{s}$, $\bar{d}_t=1~\mathrm{mm}$, $\xi=10^{-8}~\mathrm{m}$, and $\sigma = 0.1$. Medium size is $120\times120$.}
    \label{fig:45DegAngle}
\end{figure}

Finally, we examine the case where the particles are initially placed at a $45^{\circ}$ angle from each other. We expect that in the case where two particles start right above one another, they quickly move to a configuration where they are at a $45^{\circ}$ angle from one another due to the dipolar flow perturbation generated by a single particle. Fig.~\ref{fig:45DegAngle} confirms this expectation; the two different curves coincide with one another almost perfectly. \\

\begin{figure}[h]
     \centering
     \includegraphics[width=1\linewidth, trim = 100 225 105 250, clip=true]{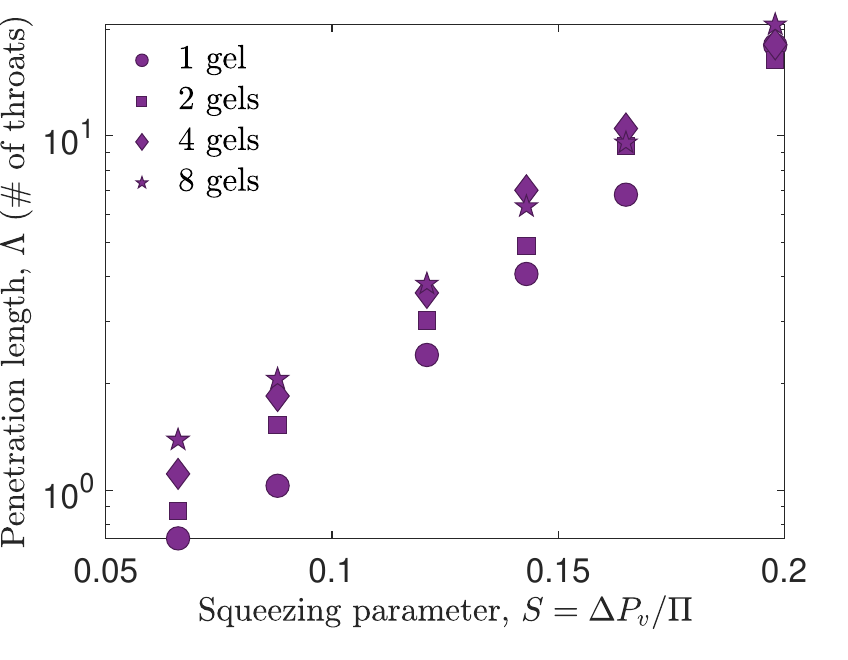}
     \caption{Mean penetration length $\Lambda$ as a function of $S$ for 2, 4, and 8 particles. All simulations use $\mu = 10^{-3}~\mathrm{Pa}\cdot\mathrm{s}$, $\bar{d}_t=1~\mathrm{mm}$, $\xi=10^{-8}~\mathrm{m}$, and $\sigma = 0.1$. Medium size is $120\times120$.}
     \label{fig:MultipleGels}
\end{figure}

\textbf{\emph{Multiple particles.}} In this section, we examine the effect of doubling and quadrupling the number of hydrogel particles in the system (as shown in Fig.~\ref{fig:MultipleGels}). 
In each case, the particles are initially separated in the transverse direction by one throat (the minimum possible separation distance). As shown by the data, the penetration length increases as the number of particles increases. 
However, this effect eventually saturates, as shown by the fact that the 4 and 8 particle curves begin to overlap with one another.\\

\textbf{\emph{Dipolar flow perturbation for two particles.}} In the case of two particles with minimal separation, the dipolar perturbation to the flow generated by one particle creates an accelerated flow field in the vicinity of the other one, thus enhancing its progress through the medium. Here, we have displayed a series of screenshots, where initially the top particle lags behind (Panel A), catches up to the second particle (B), and then the two particles proceed to exit the medium together (C and D). A time-lapse movie showing this progression is given by Supporting Video 1.


\begin{figure}[h]
    \centering
    \includegraphics[width=1\linewidth,trim = 25 65 375 65, clip=true]{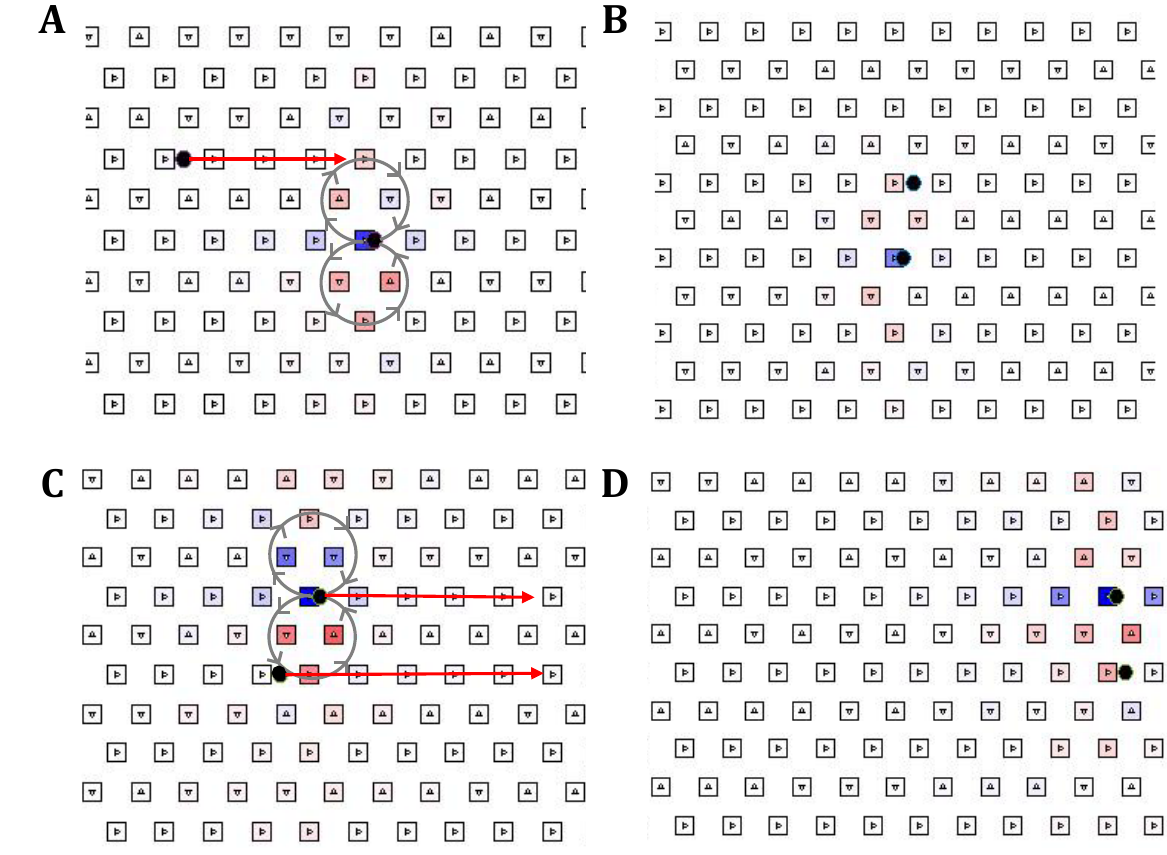}
    \caption{Series of snapshots illustrating cooperative transport of two particles in a porous medium with $\sigma=0.1$}
    \label{fig:2GelFlow}
\end{figure}

\providecommand{\noopsort}[1]{}\providecommand{\singleletter}[1]{#1}%

\end{document}